\documentclass[apj]{emulateapj}

\usepackage{graphicx, graphics}
\usepackage{amsmath,amssymb}

\received{}
\revised{}
\accepted{}

\shorttitle{X-ray Absorbers Detected in the Galactic X-ray Binaries}
\shortauthors{Luo \& Fang}

\def\fexxv{Fe\,{\sc xxv}\ }
\def\fexxvi{Fe\,{\sc xxvi}\ }

\def\chandra{{\sl Chandra\ }}
\def\xmm{{\sl XMM}-Newton\ }

\begin{document}

\title{On the Origin of Highly Ionized X-ray Absorbers Detected in the Galactic X-ray Binaries}

\author{Yang~Luo\altaffilmark{1} and Taotao~Fang\altaffilmark{1,2}}

\altaffiltext{1}{Department of Astronomy and Institute of Theoretical Physics and
  Astrophysics, Xiamen University, Xiamen, Fujian 361005, China; fangt@xmu.edu.cn} 
\altaffiltext{2}{Department of Physics \& Astronomy, 4129 Frederick Reines Hall,
  University of California, Irvine, CA 92697} 
\begin{abstract}

X-ray observations of the Galactic X-ray binaries (XRB) revealed numerous highly ionized metal absorption lines. However, it is unclear whether such lines are produced by the hot interstellar medium (ISM) or the circumstellar medium (CSM) intrinsic to the binaries. Here we present a {\sl Chandra} X-ray absorption line study of 28 observations on 12 X-ray binaries, with a focus on the \ion{Ne}{9} and \ion{Fe}{17} lines. We report the first detections of these lines in a significant amount of observations. We do not find significant dependence of the line equivalent width on the distance of the XRBs, but we do see weak dependence on the source X-ray luminosity. We also find two out of twelve selected targets show strong temporal variation of the \ion{Ne}{9} absorbers. While the line ratio between the two ion species suggests a temperature consistent with the previous predictions of the ISM, comparing with two theoretical models of the ISM shows the observed column densities are significantly higher than predictions. On the other hand, photoionzation by the XRBs provides reasonably good fit to the data. Our findings suggest that a significant fraction of these X-ray absorbers may originate in the hot gas intrinsic to the X-ray binaries, and that the ISM makes small, if not negligible, contribution. We briefly discuss the implications to the study of the Milky Way hot gas content.

\end{abstract}


\section{Introduction}

Since the launch of the {\sl Chandra} and {\sl XMM-Newton} X-ray Observatories, high resolution observations have been performed on a large number of Galactic X-ray binaries (XRBs) (for a review of the early results, see \citealp{Paerels2003}). One of the most distinguishing features is the presence of numerous X-ray absorption lines produced by almost all astrophysically important elements, and at ionization stages ranging from neutral to Hydrogen-like. Some of these features were also detected in the X-ray spectra of active galactic nuclei at $z=0$. However, the current study of these intervening absorbers is hindered by one crucial problem: We do not know where they are located. Past studies always focused on two possibilities that the intervening gas can be associated with either the interstellar medium (ISM) or the circumstellar medium (CSM) intrinsic to the XRB. Clearly, any investigation of the ISM or the CSM using the X-ray absorption lines must address this issue first.

While most previous studies assumed either an ISM or a CSM origin, a few attempts, based on a very small number of targets, have been made to resolve this issue. For instance, \citet{miller2004} argued that the lines detected in GX~339-4 are intrinsic to the source because the line width and shifting are consistent with a disk wind picture, and they did not detect similar absorption lines in a nearby XRB. However, \citet{yao2009} and \citet{cabot2013} concluded that the absorption lines seen in Cyg X-2 are produced by the ISM because the lines are narrow, show no time variability, and the wavelengths are consistent with the reference ``rest-frame wavelengths" (although these wavelengths were determined by measurements of other XRB spectra).

So far, what have we learned about these X-ray absorbers? First, some must be intrinsic to the XRBs. For example, highly ionized \fexxv and \fexxvi have been detected in some XRBs \citep[e.g.,][]{ueda2004} but not in the background AGN spectra at $z=0$ \citep[e.g.,][]{fang2003,nicastro2002,rasmussen2003,williams2005,williams2006}. If they originate in the ISM, it would be difficult to explain why they are not presented in the AGN spectra. Some neutral and low ionization metal species could be located in the ISM (see, e.g., \citealp{liao2013}). In particular, some absorption features have been detected both in the spectra of background AGNs and in the XRBs, although it is still unclear the absorption seen in AGNs is produced by hot gas in the disk or in the distant Galactic halo. Recently, \citet{miller2009} studied the variability of neutral photoelectric absorption among a variety of XRBs and concluded that the absorption is dominated by the ISM contribution. Yet these studies are based on a small number of targets and the conclusion is limited.

In the past decade, \chandra and \xmm observed a large number of Galactic X-ray binaries, and some have been observed multiple times. These observations provide an important database to study the origin of these highly ionized metal species systematically. In this paper, we plan to address the origin of the X-ray absorbers by studying a sample of XRBs with very high quality X-ray spectra. We will first investigate the temporal behavior and spatial distribution of the metal absorption lines. We will also compare the observed line properties with those expected by theories. We focus on two ISM models of the hot gas distribution: One with a disk distribution \citep[e.g.,][]{yao2006}, and one with an extended profile, as expected for an adiabatic gas in hydrostatic equilibrium \citep[e.g.,][]{maller2004}. We will also investigate the physical conditions of the X-ray absorbers if they are located in the circumstellar environment.

The paper is organized as follows. In Section 2 we describe the selection of our sample and data reduction.  In Section 3 we examine the temporal behavior of the absorption lines, and compare the measured column densities with those predicted by theoretical models. Finally, we present the conclusions and discuss the possible origins for the X-ray absorber in Section 4.

\section{Sample selection and data reduction}

\begin{figure*}
\centering
\includegraphics[width=5in,height=3in,angle=0]{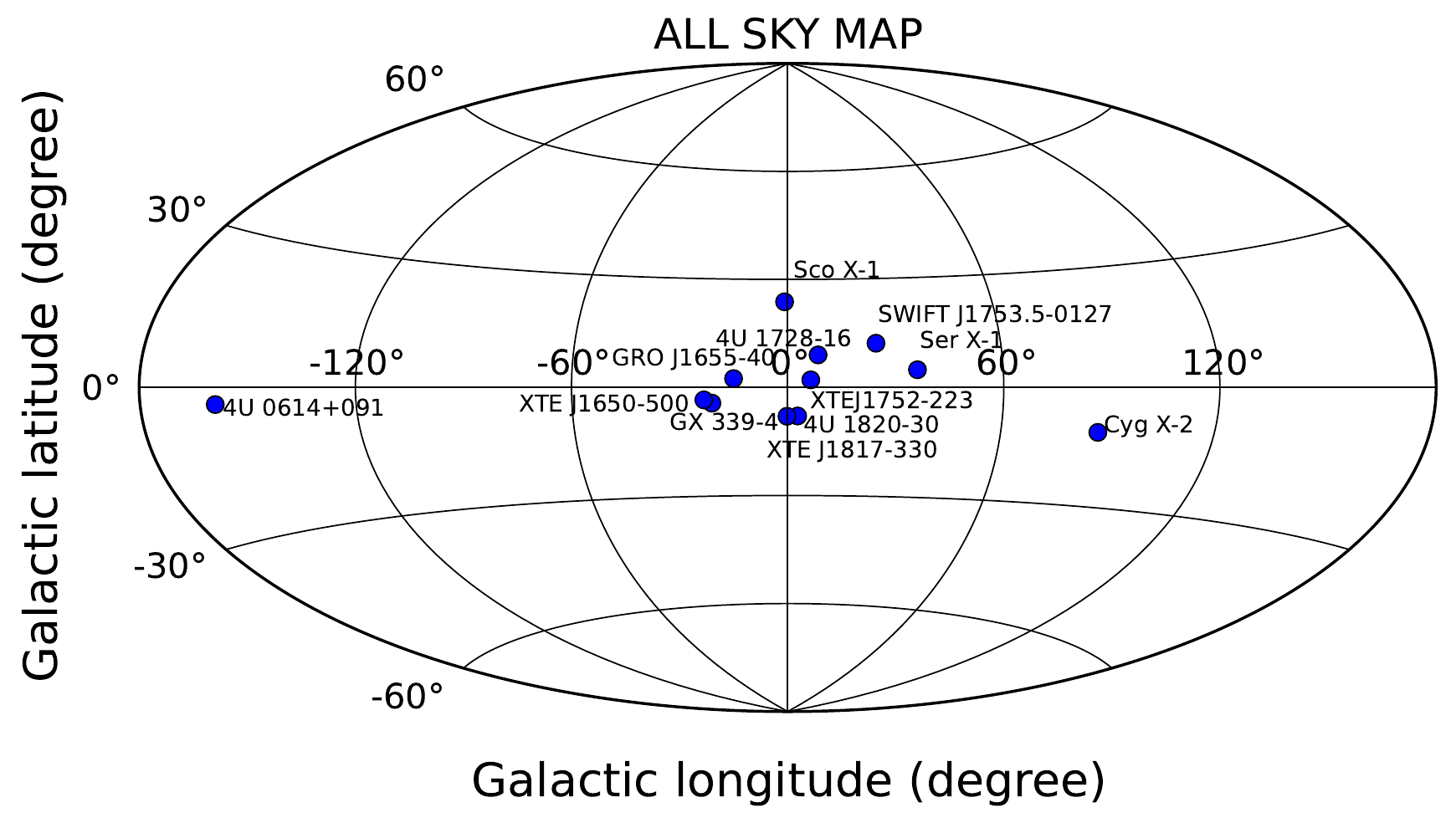}
\caption{All-sky Hammer-Aitoff projection of the 12 X-ray binaries in our sample.}
\label{fig:aitoff}
\end{figure*}

While numerous ion species have been detected in the XRB spectra, we focus on the \ion{Ne}{9} and \ion{Fe}{17} absorption lines, for several reasons. First, \ion{Fe}{17}  and the first three \ion{Ne}{9} transitions, $K_{\alpha}$, $K_{\beta}$, and $K_{\gamma}$ lines, are all located in a wavelength region (10 to 15.5 \AA)\ where the Galactic neutral hydrogen absorption becomes less important (see Table \ref{table:wavelength}). Most of the XRBs are located in the Galactic plane and the X-ray spectra suffer from strong absorption at above 15 to 20 \AA.\  Second, all three high resolution X-ray spectrometers, namely the {\sl Chandra} HETGS\footnote{High Energy Transmission Grating Spectrometer, see {\sl Chandra} Proposers' Observatory Guide at: http://cxc.harvard.edu/proposer/POG/html/.}, LETGS\footnote{Low Energy Transmission Grating Spectrometer, see {\sl Chandra} Proposers' Observatory Guide at: http://cxc.harvard.edu/proposer/POG/html/.}, and the RGS\footnote{Reflection Grating Spectrometer, see XMM-Newton Users' Handbook at: http://xmm.esac.esa.int/.} onboard \xmm have relatively high collecting area between 10 and 15 \AA.\ Combining with the fact that Neon and Iron are two of the most abundant elements in the universe, observations showed that among all the astrophysically important ion species, \ion{Fe}{17} and \ion{Ne}{9} $K_{\alpha}$ line have significantly high detection rates in the XRB spectra.

\begin{deluxetable}{lll}
\tablewidth{0pt}
\centering
\tablecaption{Absorption Line Parameters}
\tablehead{ Line & Wavelength (m\AA) & $f_{ij}$ }            
\startdata      
\ion{Ne}{9}	K$_{\alpha}$	& 13.4471	& 7.24E-01\\
\ion{Ne}{9}	K$_{\beta}$ 	& 11.5466	& 1.49E-01\\
\ion{Ne}{9} K$_{\gamma}$	& 11.0003	& 5.61E-02\\
\ion{Fe}{17}               & 15.015     & 2.31\\
\enddata
\label{table:wavelength}
\tablecomments{\ Laboratory wavelengths and oscillator strengths of \ion{Ne}{9} K$_{\alpha}$, K$_{\beta}$, K$_{\gamma}$ and also \ion{Fe}{17}. The values are adopted from \citet{verner1996}.}
\end{deluxetable}

In the 10-15.5 \AA\ waveband, absorption lines from a number of other ion species such as \ion{Mg}{11} were also detected in the spectral of several XRBs (see, e.g., \citealp{miller2004, yao2009}), and can potentially be analyzed jointly to study their nature. However, we decide to analyze \ion{Ne}{9} and \ion{Fe}{17} for the following reasons. First, unlike \ion{Ne}{9} and \ion{Fe}{17} which were detected in many XRB spectra, most other ion species are only detected in a very few targets. For example, we examined all the reported detections and found \ion{Mg}{11} was only detected in 3 XRBs, and the detection rates for other highly ionized ion species were even lower. Considering that we not only need detection, but also strong flux to constrain their property, \ion{Ne}{9} and \ion{Fe}{17} with their high detection rate certainly are the best candidates to study the global properties of these X-ray absorbers. 

Among the three high resolution X-ray spectrometers, we select the {\sl Chandra} HETGS because it has the highest spectral resolution. HETGS has two sets of grating, high energy grating (HEG) and medium energy grating (MEG). Since HEG has rather small effective area around the \ion{Ne}{9} line region, we focus on the date obtained with MEG. While the HETGS has significantly less collecting area than those of the LETGS and RGS, we prefer the HETGS for the following reasons. First, very few XRBs were observed with the LETGS. A search over Chandra archive indicated that only 15 observations on the Galactic XRBs were preformed with LETGS, only two targets were observed twice and none was observed more than twice. Second, RGS 1 has a bad CCD between 10 --- 15 \AA\ so RGS 2 is the only unit that covers this waveband. The resolution of the RGS 2 is considerably worse than those of MEG, and significantly limits the constraining power of parameter space. However, to ensure the data quality, we select the observations with at least 100 photons per 0.025 \AA\ around the \ion{Ne}{9} $K_{\alpha}$ line region. This bin size is about the full width of half maximum (FWHM) of the MEG. 

In total,  28 Chandra grating observations of twelve XRBs were selected. Among the twelve targets, eight have multiple observations and can be used for temporal analysis. In Table \ref{table:targets} we list the source name, distance, observational ID, observation date, and exposure time, in columns 1, 2, 3, 4 and 5, respectively. 


All data were reprocessed with the standard CIAO software (ver.4.4) and CALDB (ver.4.4.8). A grating RMF (gRMF) appropriate for spectral analysis of grating observations was generated by tool {\sl mkgrmf}. We calculated the auxiliary response function (ARFs) by running script {\sl fullgarf} for the first-order spectrum. Finally, the plus and minus first-order spectra were co-added using the tool {\sl add\_grating\_order}.

\section{Analysis and Results}

Segment spectra were modeled between 10 \AA\ to 15.5 \AA\ with the Galactic neutral absorption plus a power-law model in XSPEC ver 12.7 \citep{Arnaud1996}. We used a fixed ISM abundance adopted from \citet{Wilms2000} and assumed collisional ionization equilibrium for the X-ray absorbing gas. The Galactic neutral hydrogen column densities were adopted from \citet{dickeyLockman1990} and were set to be frozen in our spectra fit. Absorption lines were fitted with a Voigt line profile model. We briefly summarize its parameters here, but refer the readers to \citet{buote2009} for more details. This line model has three free parameters: Column density, Doppler parameter b, and the central wavelength shift. In our joint fit we tied together these three parameters for all three neon lines. We obtained the line equivalent width (EW) or its upper limit in case of non-detection. To obtain the EW upper limits, we assumed a velocity width of b $\sim$ 150 km s$^{-1}$. The quoted errors correspond to a 90\% confidence level unless stated otherwise. For \ion{Ne}{9}, we performed a jointly fit of the K$_{\alpha}$, K$_{\beta}$ and K$_{\gamma}$ transitions to determine the properties of the absorber. Table \ref{table:result} lists the line properties determined from the spectral fitting. Columns 3, 4, and 5 are the line EWs of \ion{Ne}{9} K$_{\alpha}$, K$_{\beta}$ and K$_{\gamma}$ transitions, respectively. Column 6 is the significance of the  K$_{\alpha}$ line. Columns 8 and 9 show the \ion{Fe}{17} line EW and significance, and column 7 and 10 are comments, respectively. Some line detections have been reported before (see, e.g., \citealp{juett2006,yao2005,cackett2008,miller2004}).

\begin{deluxetable}{lllll}
\tablewidth{0pt}
\tiny
\tablecaption{TARGETS}
\tablehead{ Source  & Distance &ObsID& Obs Time & Exp. \\
& (kpc) & & & (ksec) }
\startdata  
4U 0614+09	&	2.2	$\pm$	0.8	$^{[1]}$ &	10759	&	24-Jan-09	&	60.3	\\
	&				&	10857	&	21-Jan-09	&	58.7	\\
Sco X-1	&	2.8	$\pm$	0.3	$^{[8]}$ &	3505	&	21-Jul-03	&	16.1	 \\
4U 1728-16	&	10			$^{[2]}$ &	11072	&	13-Jul-10	&	98.0	\\
	&				&	703	&	22-Aug-00	&	21.2	\\
4U 1820-30	&	7.6	$\pm$	0.4	$^{[3]}$ &	6633	&	12-Aug-06	&	25.2	\\
	&				&	6634	&	20-Oct-06	&	25.1	\\
	&				&	7032	&	05-Nov-06	&	46.3	\\
Cyg X-2	&	13.4	$\pm$	1.9	$^{[4]}$ &	1016	&	12-Aug-01	&	15.1	\\
	&				&	10881	&	12-May-09	&	66.6	\\
	&				&	1102	&	23-Sep-99	&	29.0	\\
	&				&	8170	&	25-Aug-07	&	77.5	\\
	&				&	8599	&	23-Aug-07	&	70.6	\\
GRO J1655-40	&	3.2	$\pm$	0.2	$^{[5]}$ &	5460	&	12-Mar-05	&	14.9	\\
	&				&	5461	&	01-Apr-05	&	26.2	\\
GX 339-4	&	10	$\pm$	4.0	$^{[6][7]}$ &	4420	&	17-Mar-03	&	76.2	\\
	&				&	4569	&	22-Aug-04	&	50.1	\\
	&				&	4570	&	04-Oct-04	&	50.2	\\
	&				&	4571	&	28-Oct-04	&	51.2	\\
Ser X-1	&	11.1	$\pm$	1.6	$^{[9]}$&	700	&	13-Jun-00	&	78.1	\\
SWIFT J1753.5-0127	&	5.4	$\pm$	2.5	$^{[10]}$ &	14428	&	03-May-12	&	20.0	\\
XTE J1650-500	&	2.6	$\pm$	0.7	$^{[11]}$ &	2699	&	05-Oct-01	&	19.6	\\
	&				&	2700	&	29-Oct-01	&	28.5	\\
XTE J1817-330	&	8			$^{[12]}$ &	6615	&	13-Feb-06	&	18.1	\\
	&				&	6616	&	24-Feb-06	&	29.8	\\
	&				&	6617	&	15-Mar-06	&	47.3	\\
	&				&	6618	&	22-May-06	&	51.4	\\
XTE J1752-223 	&	3.5	$\pm$	0.4	$^{[13]}$ &	10070	&	08-Feb-10	&	19.8	\\
\enddata
\label{table:targets}
\tablecomments{Selected sample of X-ray binaries and their observation information. See the text for detailed explanation of each column. The distance for each XRB is also shown. However many of the X-ray binaries do not have well-determined distances,the most constrained distance error bar is also shown. For 4U 1728-16 and XTE J1817-330, as the errors could not be constrained, we simply assume the distances as 10 Kpc and 8 Kpc, respectively. 
[1] \citet{Paerels2001};
[2] \citet{Savolainen2009};
[3] \citet{Kuulkers2003}
[4] \citet{Smale1998}
[5] \citet{Jonker2004}
[6] \citet{Hynes2004}
[7] \citet{Zdziarski2004}
[8] \citet{Bradshaw1999}
[9] \citet{Galloway2008}
[10] \citet{Zurita2008}
[11] \citet{Homan2006}
[12] \citet{Sala2007}
[13] \citet{Shaposhnikov2010}.
}
\end{deluxetable}

Among all XRBs, the strong \ion{Ne}{9} K$_{\alpha}$ absorption lines near 13.447 \AA\ are clearly visible with good S/N spectra, except in 4U 0614+09 and Sco X-1. \ion{Fe}{17} was also detected in 9 observations (for 4 targets) at more than the $3\sigma$ level. We performed Monte-Carlo simulation to estimate the errors on the line equivalent width. The results are summarized in Table \ref{table:result}. For the non-detections we estimated their 90\% upper limits. For the ObsID 5461 of GRO J1655-40, due to the blending of strong absorption lines from ions other than \ion{Ne}{9} in the \ion{Ne}{9} K$_{\beta}$ and K$_{\gamma}$ regions, we only fitted the \ion{Ne}{9} K$_{\alpha}$ line. For this observation and the ObsID 2699 of XTE~J1650-500, our results are significantly different from those of \citet{miller2008}; for the rest observations, our best-fit values and errors are in general consistent with previous estimates.

\begin{deluxetable*}{lllllllllll}
\tablewidth{0pt}
\tabletypesize{\scriptsize}
\tablecaption{Absorption Lines Properties}
\tablehead{Source &ObsID & \multicolumn{5}{c}{NeIX} & \colhead{} & \multicolumn{3}{c}{FeXVII} \\
\cline{3-7} \cline{9-11} \\
& & EW K$_{\alpha}$&EW K$_{\beta}$&EW K$_{\gamma}$ & S/N & Comments & \colhead{} & EW & S/N & Comments\\
& & (m\AA) & (m\AA)& (m\AA)& ($\sigma$) & & & (m\AA)& ($\sigma$) & }            
\startdata                    
4U 0614+09	&	10759	&	$<$ 3.23			&				&				&		&	...	&    &	$<$ 6.46			&		&	...	\\
	&	10857	&	$<$ 4.10			&				&				&		&	...	&	   &	$<$ 5.22			&		&	...	\\
Sco X-1	&	3505 &	$<$ 4.03			&				&				&		&	... &    &	$<$ 3.23 & & ...\\
4U 1728-16	&	11072	&	5.19	$\pm$	2.30	&	1.00	$\pm$	0.47	&	0.35	$\pm$	0.16	&	6.1	&	this work	&	   &	4.50	$\pm$	4.90	&	2.5	&	this work	\\
	&	703	&	5.91	$\pm$	4.00	&	1.38	$\pm$	1.10	&	0.48	$\pm$	0.39	&	4.0	&	6	$\pm$	3$^{[a]}$	    &	   &	6.36	$\pm$	4.88	&	3.5	&	this work	\\
	&              &                                        &                                        &                                        &              &      4.37	$\pm$	3.2$^{[b]}$    &               &                             	&              &                        \\
4U 1820-30	&	6633	&	4.25	$\pm$	2.48	&	1.39	$\pm$	1.42	&	0.55	$\pm$	0.80	&	4.6	&	4$\pm$2.44$^{[c]}$	&	   &	5.26	$\pm$	3.87	&	3.7	&	5.1$\pm$2.2$^{[f]}$	\\
	&	6634	&	5.87	$\pm$	3.44	&	0.99	$\pm$	0.53	&	0.34	$\pm$	0.18	&	4.6	&	5$\pm$2.17$^{[c]}$	&	   &	3.28	$\pm$	3.80	&	2.3	&	5.1$\pm$2.2$^{[f]}$	\\
	&	7032	&	5.25	$\pm$	1.72	&	1.54	$\pm$	1.05	&	0.57	$\pm$	0.50	&	8.3	&	this work	&	   &		3.90	$\pm$	2.87	&	3.7	&	this work	\\
Cyg X-2	&	1016	&	7.09	$\pm$	4.80	&	1.14	$\pm$	0.83	&	0.39	$\pm$	0.28	&	3.9	&	3.66	$\pm$	2.2$^{[b]}$ &	   &	$<$ 4.11			&		&	$<$1.0$^{[g]}$	\\
	&              &                                        &                                        &                                        &              &      2.7$\pm$0.2$^{[g]}$    &               &                             	&              &                        \\
	&	10881	&	4.80	$\pm$	1.81	&	0.79	$\pm$	0.30	&	0.27	$\pm$	0.10	&	7.3	&	this work	&	   &	$<$ 2.35			&		&	...	\\
	&	1102	&	3.54	$\pm$	4.23	&	0.79	$\pm$	0.90	&	0.27	$\pm$	0.31	&	2.4	&	$<$ 5$^{[a]}$	&	   &	1.39	$\pm$	2.60	&	1.4	&	this work	\\
	&	8170	&	3.33	$\pm$	0.97	&	0.76	$\pm$	0.25	&	0.26	$\pm$	0.42	&	9.3	&	this work	&	   &	1.98	$\pm$	2.12	&	2.5	&	this work	\\
	&	8599	&	3.66	$\pm$	1.46	&	0.59	$\pm$	0.23	&	0.20	$\pm$	0.08	&	6.8	&	this work	&	   &	$<$ 1.59			&		&	...	\\
GRO J1655-40	&	5460	&	2.65	$\pm$	2.99	&	0.53	$\pm$	0.57	&	0.19	$\pm$	0.23	&	2.5	&	this work	&	   &	4.14	$\pm$	4.16	&	2.7	&	this work	\\
	&	5461	&	2.96	$\pm$	2.93	&		-		&		-		&	2.7	&	22 $\pm$13.6$^{[d]}$	&	   &	$<$ 2.03			&		&	...	\\
GX 339-4	&	4420	&	12.27	$\pm$	1.99	&	3.79	$\pm$	0.89	&	1.48	$\pm$	0.45	&	16.7	&	11.6	$\pm$	1.3$^{[a]}$	&	   &	$<$ 6.94			&		&	$<$1.6$\pm$1.6$^{[e]}$	\\
	&              &                                        &                                        &                                        &              &      17$\pm$2.7$^{[e]}$  &               &                             	&              &                        \\
	&	4569	&	9.35	$\pm$	1.48	&	3.64	$\pm$	0.72	&	1.55	$\pm$	0.45	&	17.1	&	this work	&	   &	3.00	$\pm$	2.87	&	2.8	&	this work	\\
	&	4570	&	8.16	$\pm$	1.36	&	3.29	$\pm$	0.66	&	1.44	$\pm$	0.44	&	16.2	&	this work	&	   &	2.49	$\pm$	2.06	&	3.3	&	this work	\\
	&	4571	&	8.42	$\pm$	1.40	&	3.43	$\pm$	0.66	&	1.50	$\pm$	0.44	&	16.3	&	this work	&	   &	2.40	$\pm$	2.25	&	2.9	&	this work	\\
Ser X-1	&	700	&	8.10	$\pm$	2.89	&	1.88	$\pm$	0.86	&	0.68	$\pm$	0.35	&	7.7	&	7.7	$\pm$	1.4$^{[a]}$&	  &	$<$ 5.41			&		&	...	\\
	&              &                                        &                                        &                                        &              &     6.13	$\pm$	0.72$^{[b]}$      &       &               	&              &                        \\
 J1753.5-0127	&	14428	&	3.29	$\pm$	4.03	&	0.60	$\pm$	0.51	&	0.34	$\pm$	0.62	&	2.4	&	this work	&	  &	$<$ 9.39			&		&	...	\\
XTE J1650-500	&	2699	&	3.63	$\pm$	2.02	&	1.09	$\pm$	0.78	&	0.51	$\pm$	0.61	&	4.9	&	0.7$\pm$2.98$^{[e]}$	&	  &	$<$ 5.69			&		&	...	\\
	&	2700	&	4.85	$\pm$	2.22	&	0.93	$\pm$	0.45	&	0.33	$\pm$	0.16	&	5.8	&	3.9 $\pm$ 2.4$^{[e]}$	&	  &	$<$ 4.55			&		&	...	\\
XTE J1817-330	&	6615	&	6.22	$\pm$	1.79	&	1.00	$\pm$	0.29	&	0.34	$\pm$	0.10	&	9.4	&	this work	&	  &	1.84	$\pm$	0.92	&	5.4	&	this work	\\
	&	6616	&	6.14	$\pm$	1.33	&	1.03	$\pm$	0.21	&	0.36	$\pm$	0.07	&	12.5	&	this work	&	  &	2.24	$\pm$	0.97	&	6.2	&	this work	\\
	&	6617	&	5.86	$\pm$	1.09	&	1.10	$\pm$	0.22	&	0.39	$\pm$	0.08	&	14.5	&	this work	&	  &	1.51	$\pm$	1.31	&	3.1	&	this work	\\
	&	6618	&	5.50	$\pm$	1.43	&	2.20	$\pm$	0.87	&	0.98	$\pm$	0.61	&	10.4	&	this work	&	  &	3.32	$\pm$	2.22	&	4.0	&	this work	\\
XTE J1752-223 	&	10070	&	3.71	$\pm$	2.93	&	0.95	$\pm$	0.81	&	0.34	$\pm$	0.33	&	3.4	&	this work	&	  &	$<$ 4.35			&		&	...	\\
\enddata
\label{table:result}
\tablecomments{ We show the measured Ne XI and Fe XVII equivalent width for each detection with their errors derived from Monte Carlo simulations.  Significances were calculated by dividing the K$_{\alpha}$ EW by its 1$\sigma$ error. Since most of the observation have been reported before, we convert the confidence lever of prior measured line equivalent width to 90\% and list the values for a comparison.
[a] \citet{juett2006}
[b] \citet{yao2005}
[c] \citet{cackett2008}
[d] \citet{miller2008}
[e] \citet{miller2004}
[f] \citet{yao2006}
[g] \citet{yao2009}
}
\end{deluxetable*}

\subsection{Variability}

\begin{figure*}
\center
\includegraphics[width=.45\textwidth,height=.3\textheight,angle=0]{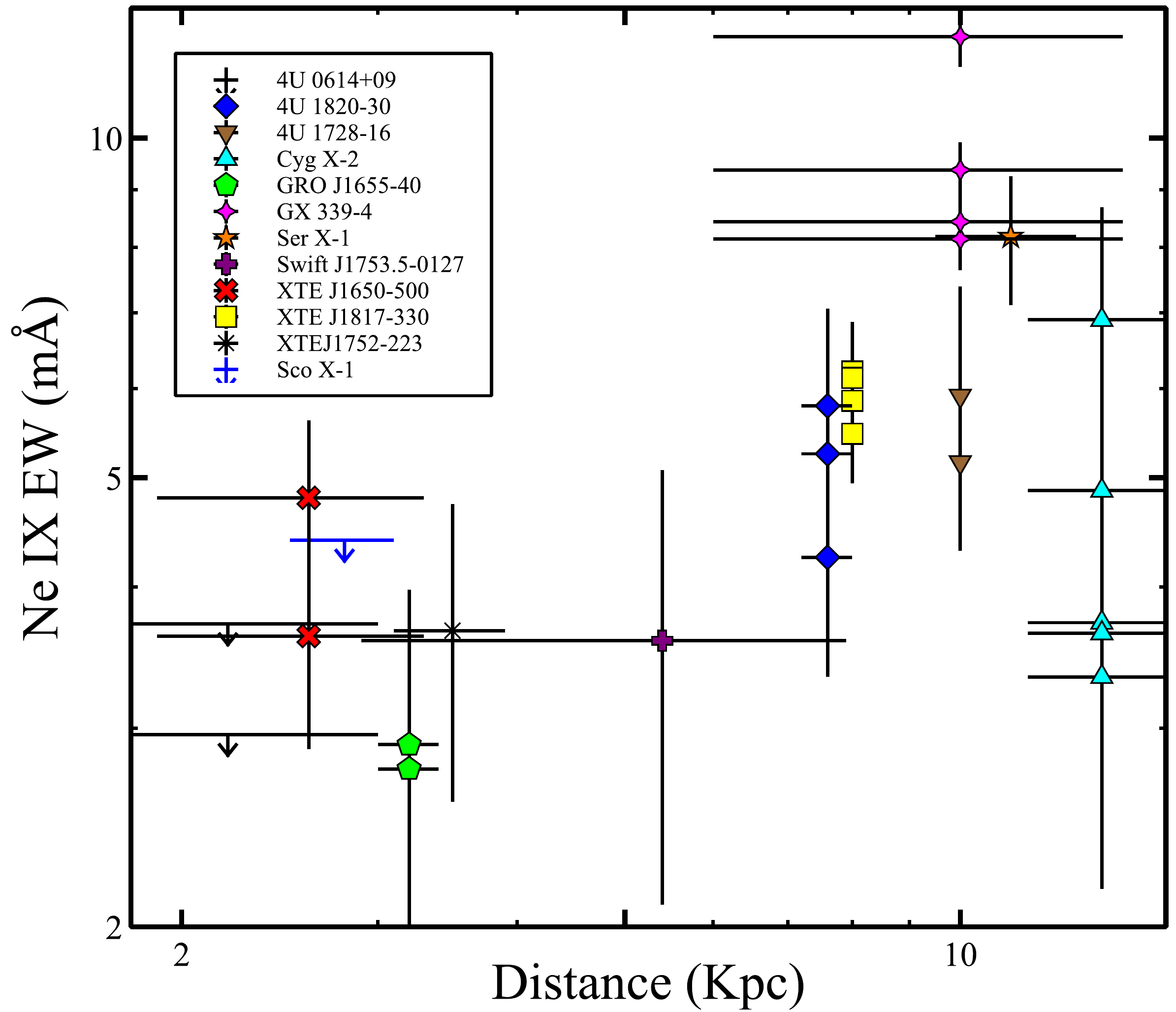}
\hskip0.5in
\includegraphics[width=.45\textwidth,height=.3\textheight,angle=0]{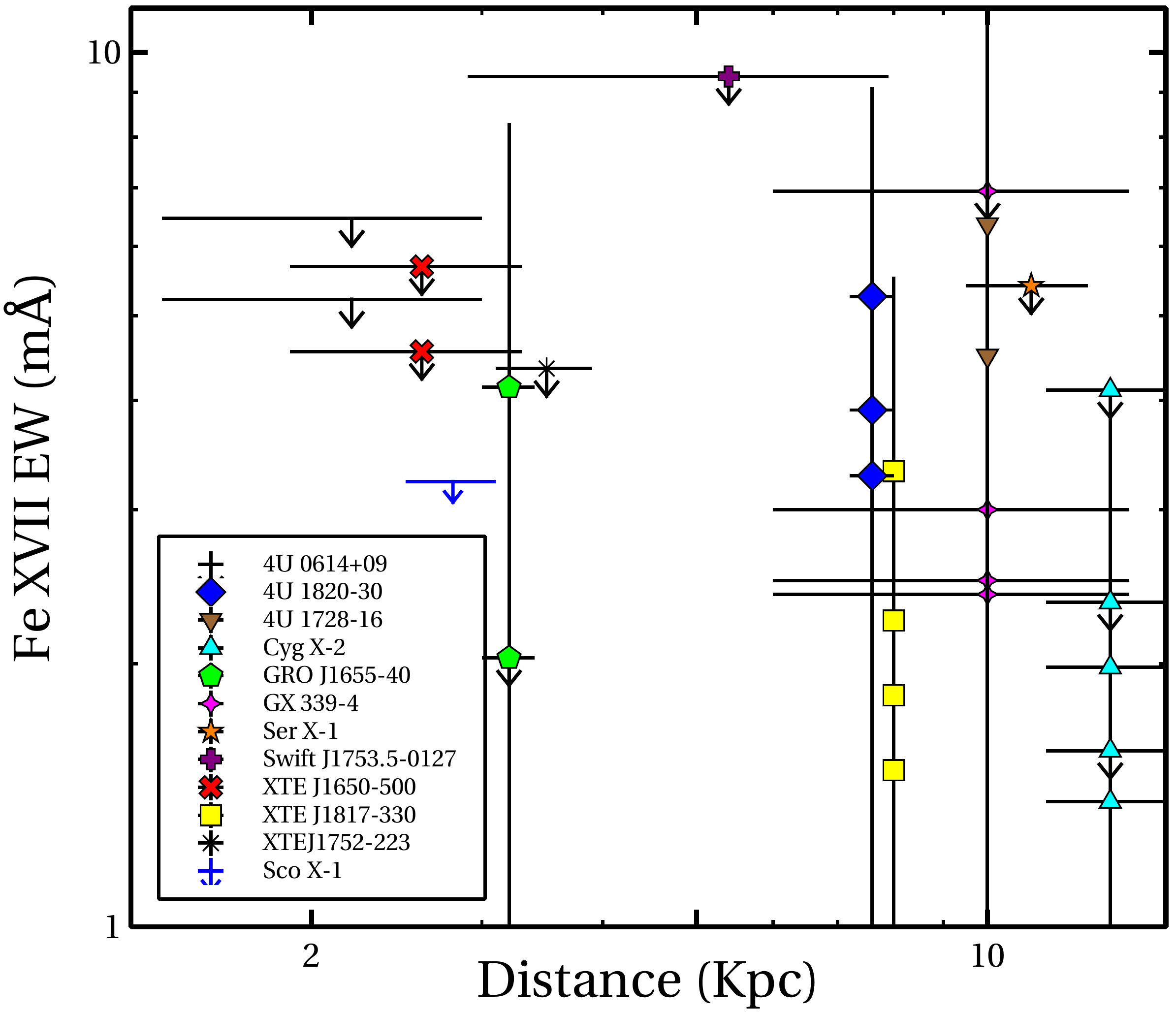}
\caption{Measured absorption line equivalent width as a function of source distance. For non-detection cases, we show their 90\% upper limits. The left panel show the relation for \ion{Ne}{9} K$_{\alpha}$ and the right panel for \ion{Fe}{17}.}
\label{fig:dis2ew}
\end{figure*}

\begin{figure}[!t]
\center
\includegraphics[width=0.45\textwidth,height=0.3\textheight,angle=0]{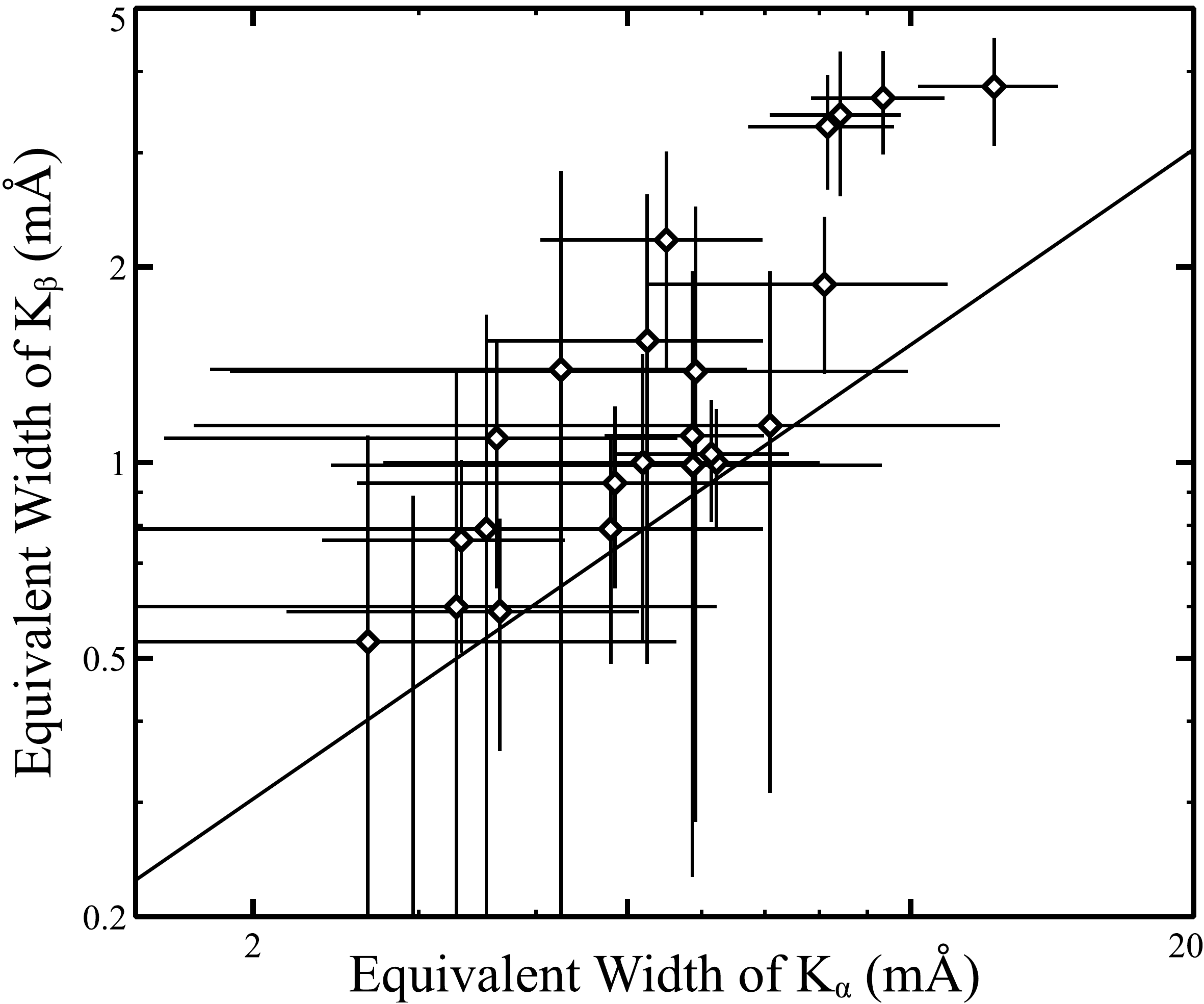}
\caption{ \ion{Ne}{9} K$_{\alpha}$ equivalent width vs. K$_{\beta}$. The solid line indicates the fixed ratio of 0.15 if the absorption is not saturated.}
\vskip0.3cm
\label{fig:ratio}
\end{figure}

We first investigate the temporal behavior of the X-ray absorption lines. An ISM-origin would imply that these absorbers should not change between different observations for the same background sources. We present our EWs for all the sample targets as function of source distance in Figure \ref{fig:dis2ew} (left panel: \ion{Ne}{9} $K_{\alpha}$; right panel: \ion{Fe}{17}). While we do see variations among targets with multiple observations, the error bars are too large to draw any reliable conclusions. We also do not detect any dependence of the line EW on the distance. In Figure \ref{fig:ratio}, we show the \ion{Ne}{9} $K_{\beta}$ line EW vs. $K_{\alpha}$ line EW. Since $\rm EW \propto f_{ij}\lambda^2$, we expect a fixed ratio of $\rm EW(K_{\beta})/EW(K_{\alpha}) \approx 0.15$ (solid line) if the line is not saturated. Since most data points are located at above the solid line, we find a significant amount of lines are saturated.

\begin{figure*}[!t]
\center
\includegraphics[width=.45\textwidth,height=.3\textheight,angle=0]{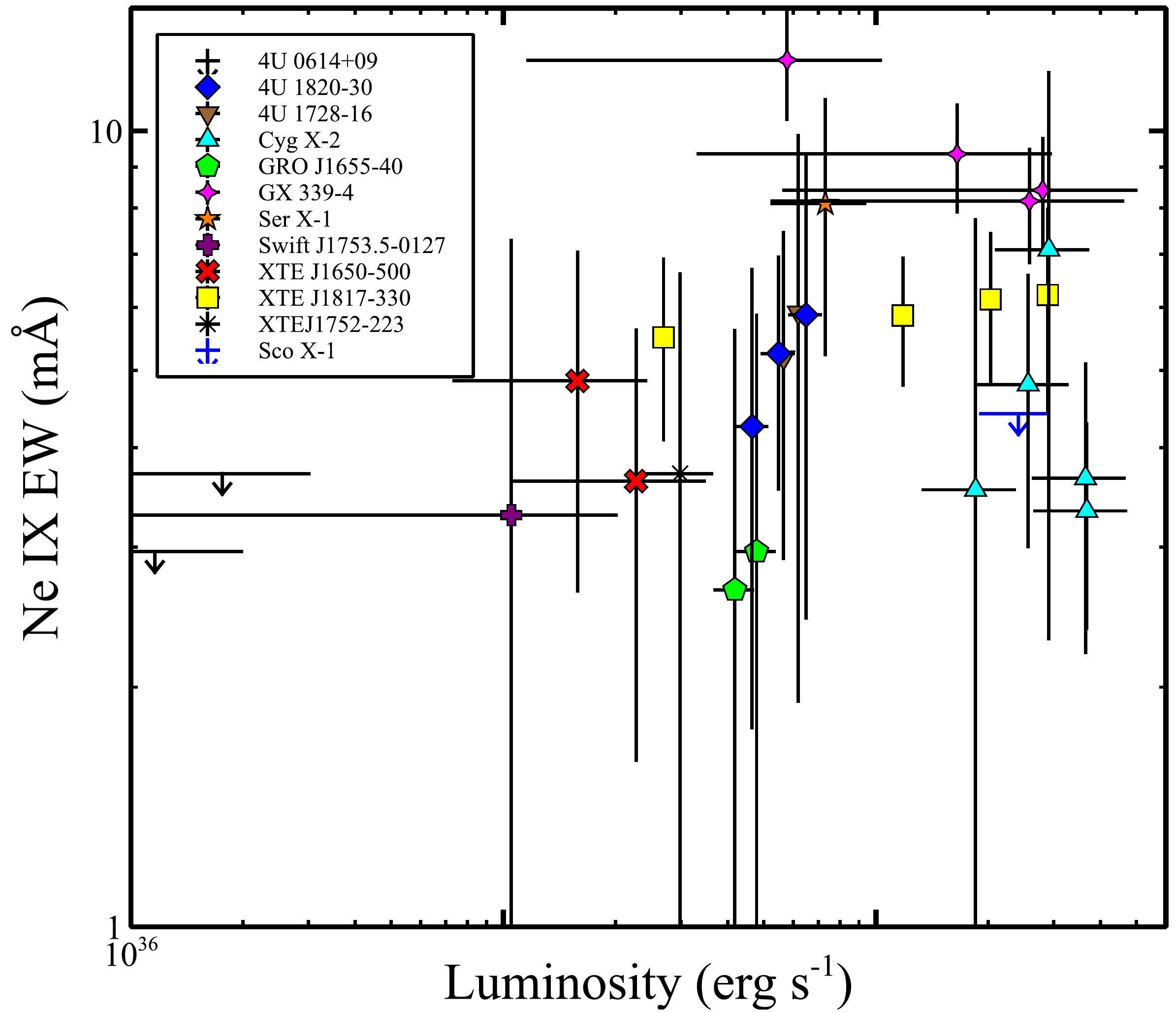}
\hskip0.5in
\includegraphics[width=.45\textwidth,height=.3\textheight,angle=0]{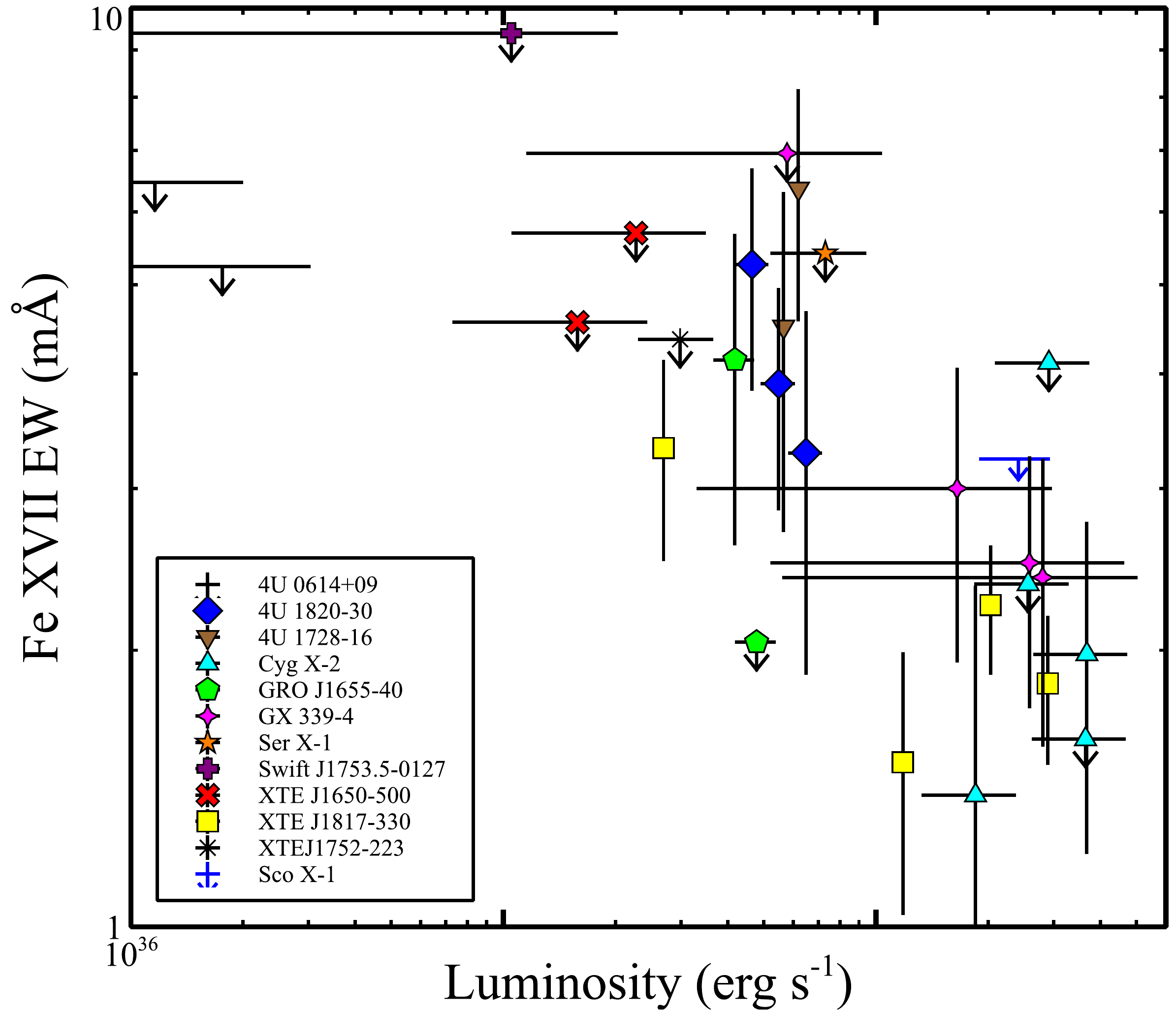}
\caption{Absorption line equivalent width of \ion{Ne}{9} $K_{\alpha}$  (left panel) and \ion{Fe}{17} (right panel) as a function of XRB X-ray luminosity between 0.6 and 6 keV.}
\vskip0.3cm
\label{fig:lum}
\end{figure*}

To investigate the line variation, we calculate the 1 $\sigma$ and 2 $\sigma$ confidence contours on \ion{Ne}{9} column density and the Doppler parameter (we focus on \ion{Ne}{9} here because this ion was detected in most observations). We would expect for the same target the parameter contours for different observations should overlap in case of constant ISM absorption. The confidence contours for all targets with detected \ion{Ne}{9} lines are displayed in Figure \ref{fig:contour}, where different colors represent different observations. The parameter contours from GX~339-4 and XTE~J1817-330 clearly show lines properties changed between observations. Our result confirmed \citet{miller2004} that the absorption detected in GX~339-4 is most likely intrinsic. For other sources, their line variabilities are less constrained. Our contour plots suggest that at least in some cases the line parameters show significant variations between observations that can not be explained by an ISM-origin of the hot gas. 

To further corroborate the connection between the absorption line and the background source, {we plot the line EW as a function of the source luminosity in Figure~\ref{fig:lum}, in which the left panel shows case for \ion{Ne}{9} $K_{\alpha}$ line and the right panel for \ion{Fe}{17}. The luminosity is calculated between 0.6 and 6 keV, the effective energy band for {\sl Chandra} MEG. The error is dominated by the uncertainty on distance. For \ion{Ne}{9}, we examined the four sources with more than two observations. Except Cyg~X-2, we do see some weak correlations for 4U~1820-30, GX~339-4, and XTE~J1817-330. The correlation is negative for  GX~339-4, and positive for the other two sources. While the detection of \ion{Fe}{17} is fewer, we do see a general trend in which the line EW becomes smaller as the source luminosity increases. Such correlation does imply links between the X-ray absorbers and the circumstellar material. We present a simply photoionization model in the next section, and refer readers to \citet{miller2004} and \citet{cackett2008} for further discussion.

We also would like to emphasize that the detection of the time variability cannot rule out the scenario that both the intrinsic absorbers and the ISM contribute to the observed absorption. Rather, such variability suggests that a significant fraction of the observed absorption should arise from regions intrinsic to the XRBs.

\subsection{Comparison with theories}

\begin{figure}[!t]
\center
\includegraphics[width=0.45\textwidth,height=0.3\textheight,angle=0]{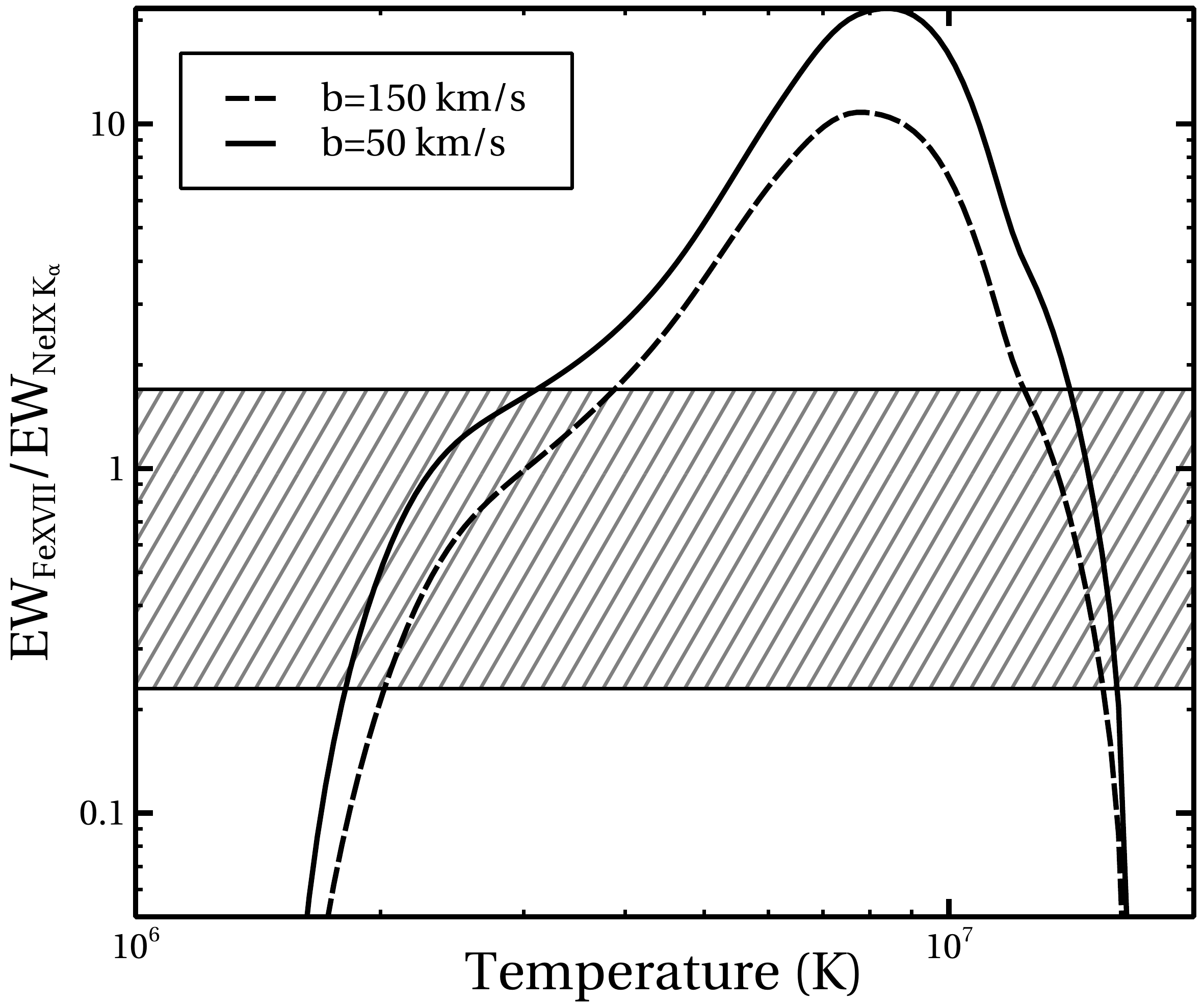}
\caption{The ratio of \ion{Fe}{17} line EW and \ion{Ne}{9} $K_{\alpha}$ as a function of temperature for collisional ionization. The line EW is calculated assuming a Doppler-$b$ parameter of 50 (solid curve) and 150 (dashed curve) $\rm km\ s^{-1}$. The shaded area indicates the region allowed by data.}
\vskip0.3cm
\label{fig:coll}
\end{figure}

\begin{figure}[!t]
\center
\includegraphics[width=0.45\textwidth,height=0.3\textheight,angle=0]{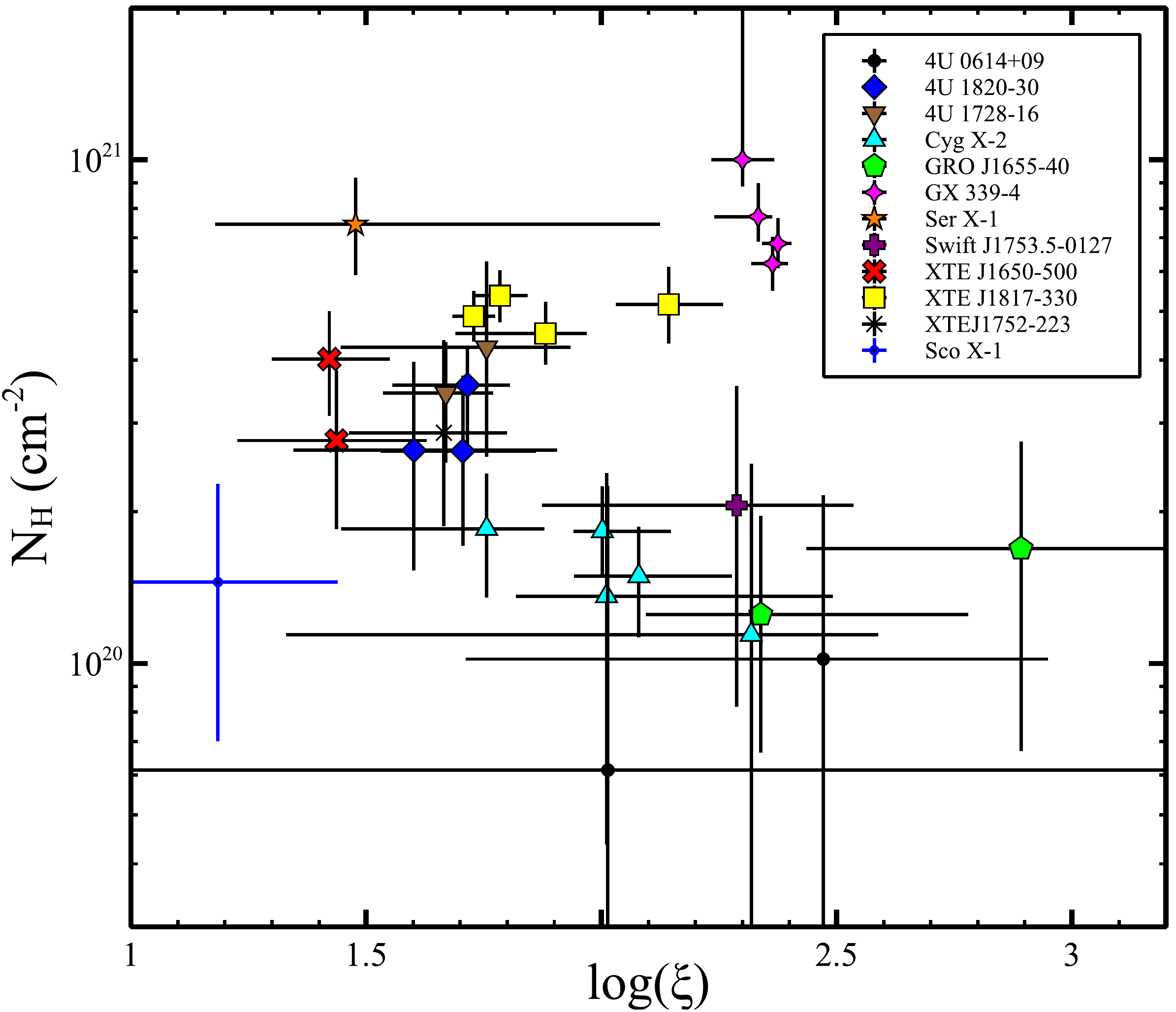}
\caption{Plot of best fitted parameter N$_{\rm H}$ vs. $\xi$ from XSTAR photoionization models.}
\vskip0.3cm
\label{fig:nh2xi}
\end{figure}

\begin{figure*}[!t]
\center
\includegraphics[width=0.8\textwidth,height=1.0\textheight,angle=0]{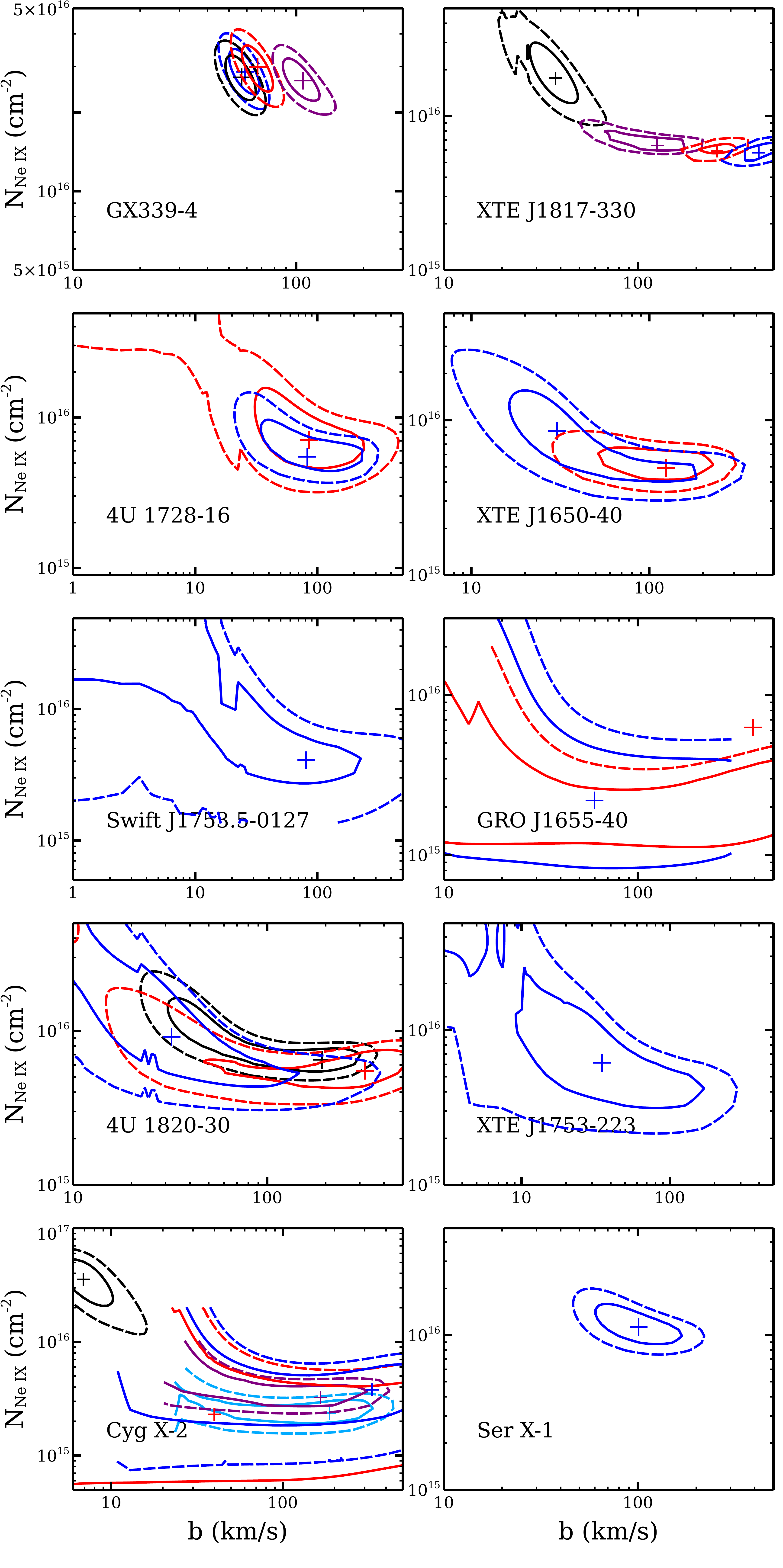}
\caption{Confidence contours representing 1$\sigma$ (solid) and 2$\sigma$ (dashed) levers on the column density and Doppler parameter b for all detected sources. The cross in each contour is the best fit value.}
\vskip0.3cm
\label{fig:contour}
\end{figure*}

\subsubsection{Collisional Ionization}

In this section, we explore the theoretical implications of our data. Collisional ionization will be the dominated ionization mechanism if the absorbers are located in the ISM. The line EW ratio between the two ion species provides an effective diagnostic tool of the ISM. In Figure~\ref{fig:coll} we plot such ratio as a function of temperature. The ionization fraction of \ion{Ne}{9} and \ion{Fe}{17} was adopted from \citet{Mazzotta1998}. We assume solar abundance and calculate the ratio for Doppler-$b$ parameter of 50 (solid curve) and 150 (dashed curve) $\rm km\ s^{-1}$ cases. The shaded area indicates the region allowed by data. We find that a temperature range of 2 --- 4 $\times 10^6$ K is predicted by data ($\sim 2\times 10^7$ K is also allowed by the ratio, but the temperature is unlikely that higher because the expected column densities of both ions would be much lower than actually detected). This temperature range is in general consistent with other measurements of the ISM temperature (see, e.g., \citealp{yao2009}). 

We also investigate the column density distribution by comparing the observations with two theoretical models of the hot gas in our Galaxy: A traditional disk-distribution model, and an extended hot-halo model based on multiphase cooling theory proposed by \citet{maller2004}. Following \citet{yao2009}, we adopt an exponential distribution and modeled the gas density and temperature distributions as:
\begin{equation}
	\rho_{\rm disk}(r) = \rho_0  \exp(\frac{-r \sin(b)}{h_{\rm \rho}  \eta}),
\end{equation}
	and
\begin{equation}
		T_{\rm disk}(r) = T_0  \exp(\frac{-r  \sin(b)}{h_{\rm T} \eta}),
\end{equation}
where $\eta$ is the volume filling factor and is assumed to be 1, $r$ is the radius from the Galactic center, $b$ is the latitude, $\rho_0$, T$_0$ is the gas density and temperature at the disk mid-plane, and h$_{\rm \rho}$ and h$_{\rm T}$ are the scale height of the hot gas density and temperature distributions, respectively. We adopt the disk model parameters from recent observations of the PKS 2155-304 sight line \citep{hagihara2010} as $\rho_0$ = 1.4$\times$10$^{-3}$ cm$^{-3}$, $T_0$ = 2.5$\times$10$^6$ K, $h_\rho$ = 2.3 kpc and $h_{\rm T}$ = 5.6 kpc. 

The extended hot halo model \citep{maller2004} predicts a hot gas density and temperature distribution as
\begin{equation}
  	\rho_{\rm mb}(r) = \rho_{\rm v} [1+\frac{3.7 R_{\rm s}}{r} \ln(1+\frac{r}{R_{\rm s}}) - \frac{3.7}{C_{\rm v}} \ln(1+C_{\rm v})]^{3/2},
\end{equation}
	and
\begin{equation}
	T_{\rm mb}(r) = T_{\rm v} [1 +\frac{3.7 R_{\rm s}}{r}\ln(1+\frac{r}{R_{\rm s}})-\frac{3.7}{C_{\rm v}}\ln(1+C_{\rm v})],
\end{equation}
Here $R_{\rm s}=21.7$ kpc is the scale radius of the dark matter halo and $C_{\rm v}=12$ is the halo concentration. The density $\rho_{\rm v}= 5.0 \times 10^{-5} cm^{-3}$ and temperature $T_{\rm v}=1.1\times10^6$ K is the gas density and temperature at the virial radius, respectively. We adopted these parameter values from \citet{maller2004} and \citet{fang2013}.

We first estimate the column density by integrating the ion number density over the sight line
\begin{equation}\label{eq:column}
	N_{\rm ion} = \int_{0}^{D} \rho(s)  f(T)  Z_{\rm ion} ds,
\end{equation}
where $\rho(s)$ is the number density along the line of sight, $D$ is the  line-of-sight distance in the direction $(l,b)$, $f(T)$ is the ionization fraction of at the expected temperature $T$, and $Z_{\rm ion}$ is the metal abundance. Note that we need to convert the model density and temperature distribution along galactic center to a direction of sight line via $r^2 = s^2+r_0^2-2 s r_0 \cos(l) \cos(b)$, where $r_0$ = 8 kpc is the distance between the Sun and the Galactic Center, and $r$ is the Galactocentric distance. Again, the ionization fractions of \ion{Ne}{9} and \ion{Fe}{17} are adopted from \citet{Mazzotta1998}, under the assumption of collisional ionization equilibrium. In what follow, we assume a $Z_{\rm ion} = 1\ Z_\odot$ for the disk model, and $0.3\ Z_\odot$ for the extended halo model, where $Z_\odot$ is the solar abundance from \citet{Wilms2000}.

\begin{figure*}[!t]
\center
\includegraphics[width=\textwidth,height=.3\textheight,angle=0]{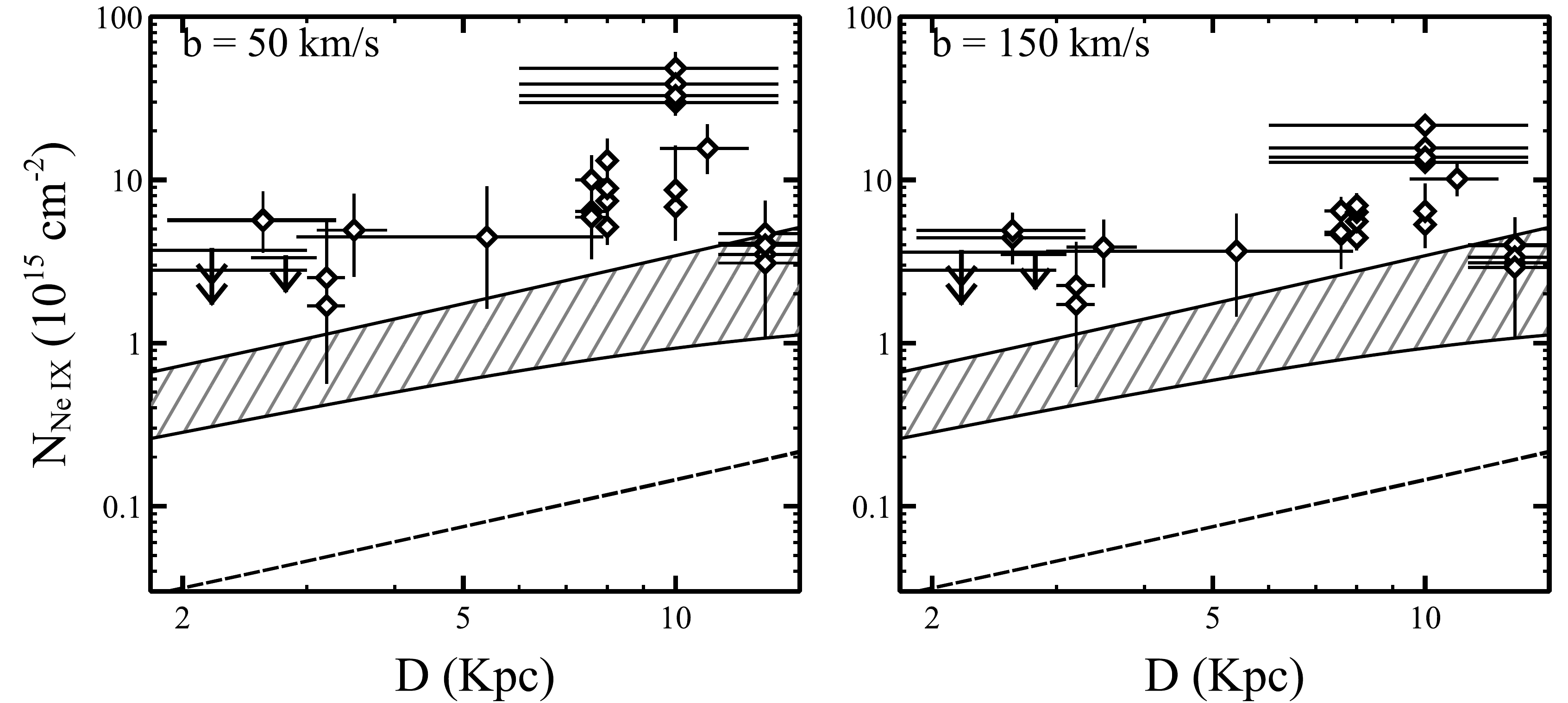}
\includegraphics[width=\textwidth,height=.3\textheight,angle=0]{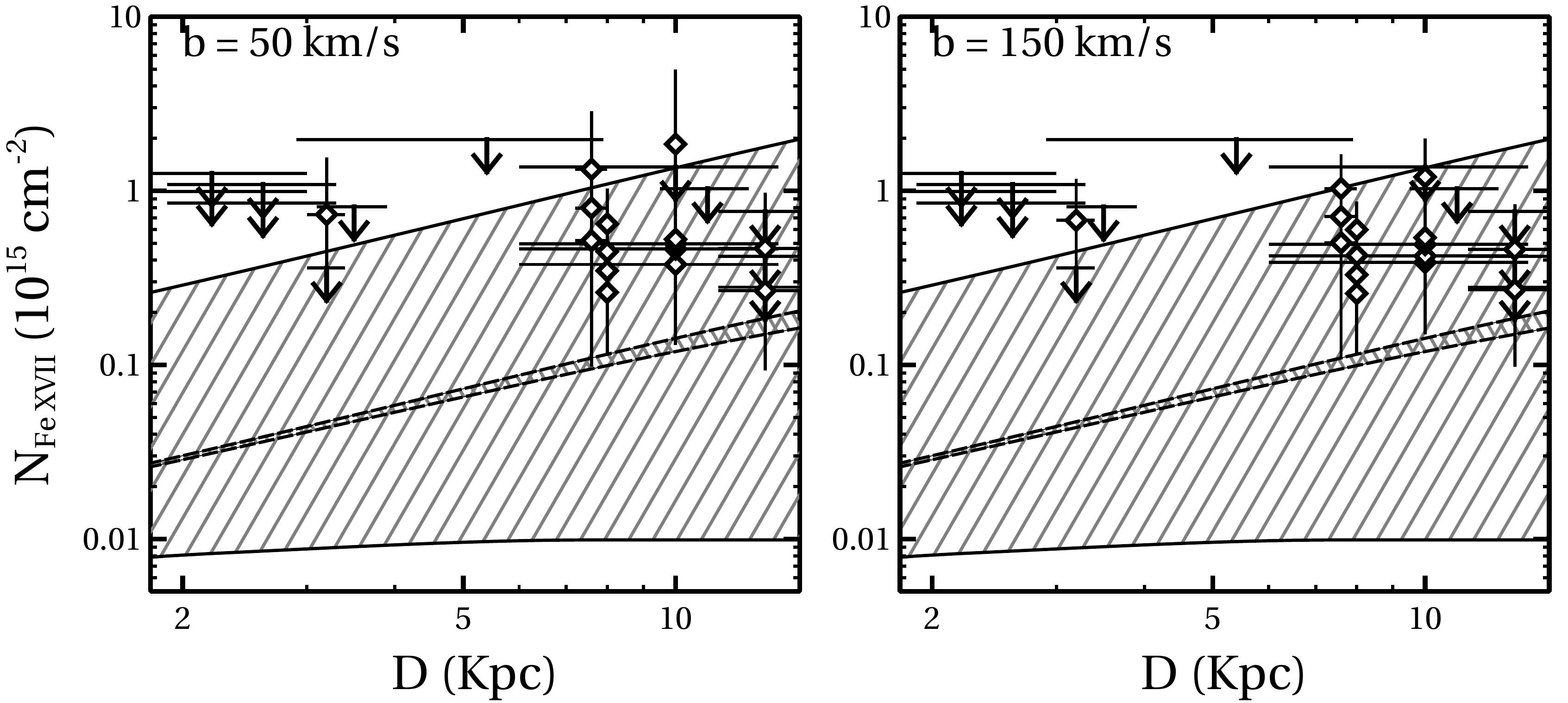}
\caption{Column density as a function of distance. The top panels show the relation for \ion{Ne}{9} in Doppler parameter b = 50 km s$^{-1}$ case (top left), and b = 150 km s$^{-1}$ case (top right). Similarly, the bottom panels show cases for \ion{Fe}{17}. In all panels, shaded area between solid lines shows the predictions from the disk model in which the upper boundary represents $(l,b) = (0,0)$ and lower boundary is $(l,b) = (0,15^{\circ})$ and area between dashed lines indicate predictions from the extended hot halo model. It is noted that the two dashed line coincidence in the top panels.}
\label{fig:dis2nh}
\vskip0.1in
\end{figure*}

The comparison of the observed column density with that expected from current theoretical ISM distributions is presented in Figure \ref{fig:dis2nh}. The top panels show the distribution for \ion{Ne}{9}, and the bottom panels for \ion{Fe}{17}. Since the column density and Doppler-$b$ parameter are correlated, we calculated the column density for all targets at a fixed $b$ of 50 (left panels) and 150 $\rm km\ s^{-1}$ (right panels). The dashed line in each panel shows the column density expected from the extended halo model. Since the density profile in this model has a rather flat core in the center, the column density distribution is insensitive to the direction of sight lines. For the disk model, we calculate two cases: $(l,b) = (0,0)$ (the lower boundary of the shaded area), and $(l,b) = (0,15^{\circ})$ (the upper boundary of the shaded area). Since most of our targets have $|b| < 15^{\circ}$, the shaded area represent the likely range of the column density distribution. Due to the lack of detections the \ion{Fe}{17} distributions (bottom two panels) cannot provide tight constraint. However, as shown in the \ion{Ne}{9} distributions (top two panels),  the column density predicted by the current ISM model provides only a small fraction to the observed XRBs absorption at less than 10-20\% level. Significant contributions have to come from sources other than the hot ISM, e.g., absorbers intrinsic to the XRBs.

\subsubsection{Photoionization}

If the X-ray absorbers are intrinsic to the XRBs, most likely photoionization will become the dominant ionization mechanism. We used the photoionization code XSTAR \citep{kallman01} to investigate the physical conditions of these absorbers. We assumed an input spectrum of a power-law with a photon index of $\Gamma$. We then created a grid of absorption line models with three grid parameters: Photon index of $\Gamma$, column density of the absorber $N_H$, and ionization parameter $\xi=L/nR^2$. Here $L$ is the X-ray luminosity, $n$ and $R$ are the density and distance (to the source). Assuming solar abundance, we folded the power-law plus the Galactic absorption model with this model spectrum calculated from XSTAR. We then fitted the observed spectra from 10.5 to 15.5 \AA\, where \ion{Ne}{9} and \ion{Fe}{17} present with XSPEC. For all fits the reduced $\chi^2$ are around 1.1, indicating our fits are reasonably good. Figure~\ref{fig:nh2xi} shows the best-fit $N_H$ vs $\xi$. For most absorbers the $N_H$ is in the range of $10^{20}$--$10^{21}\rm\ cm^{-2}$, and $\log(\xi)$ in the range of 1 -- 3. We notice in several targets, in particular Cyg~X-2, XTE~J1817-330, and to some degree GX~339-4, the physical conditions of the absorbers varied significantly between different observations.

\section{Conclusion \& Discussion}

Understanding the origin of XRB X-ray absorbers has fundamental impact on both modeling the ISM and probing the circumstellar environment of the X-ray binaries. We briefly summarize our findings here:

\begin{itemize}

\item We analyzed 28 {\sl Chandra} grating observations of 12 Galactic X-ray binaries, with a focus on the \ion{Ne}{9} $K_{\alpha}$, $K_{\beta}$, $K_{\gamma}$, and \ion{Fe}{17} absorption lines. We report the first detection of these lines in a significant amount of observations.

\item We do not find significant dependence of the observed line EW on the distance to the XRBs. Such dependence would be predicted by the ISM model of the absorbers. The ratios between the \ion{Ne}{9} $\rm K_\beta$ and $\rm K_{\alpha}$ line also suggest that they are not in the linear part of the curve-of-growth, i.e., these lines are moderately saturated. We also investigate the temporal behaviors of the absorption lines. By examining the variation of the line EW, and the contour plot of column density vs. the Doppler-$b$ parameters, we find in two XRBs, the line variation cannot be explained by an ISM-origin of the absorbers.

\item We find a weak correlation between the \ion{Ne}{9} $K_{\alpha}$ line EW and the source X-ray luminosity in three targets. We also find an overall dependence of the \ion{Fe}{17} line EW on the luminosity: The higher the luminosity, the weaker the line EW.

\item Assuming collisional ionization, the line ratio between two ion species suggests a temperature of the absorbing gas in the range of 2 --- 4$\times10^6$ K, consistent with previous estimates of the ISM. However,  we also compare the data with two theoretical models of hot gas distribution in our Galaxy: A conventional disk distribution, and an extended hot halo model. We find both models significantly underpredict the \ion{Ne}{9} column densities. 

\item We used the photoionization code XSTAR to model the absorbers, assuming there are intrinsic to the XRBs. We find in general such model provides reasonably good fit to the data. For most absorbers, their $N_H$ is in the range of $10^{20}$--$10^{21}\rm\ cm^{-2}$, and the ionization parameter $\log(\xi)$ in the range of 1 -- 3.

\end{itemize}

Based on these findings, we suggest that the ISM may make very small contribution to the total absorbing column densities, and the majority of the \ion{Ne}{9} and \ion{Fe}{17} absorbing gas arise from the vicinity of the XRBs. 

In our calculation, we assume a solar abundance in the disk model, and 0.3 solar abundance in the extended hot halo model. While the ISM may have a higher metal abundance, in order to explain the observed column densities, the abundance must be at least 5 to 10 times higher in the disk model, and more than 10 times higher in the extended hot halo model. Such abundance would be inconsistent with many other observations of the ISM. Furthermore, it would still be difficult to explain the line variation we found in several targets.

While it is interesting to perform a joint analysis of X-ray absorption and emission, we emphasize that due to different physical processes emission and absorption may probe different regions of the ISM/CSM. Since X-ray emission is proportional to the gas density square, it samples denser regions of the ISM, while absorption samples more smooth component of the ISM/CSM. Therefore in this work we restrict our investigation to the absorption study only.

Nevertheless, line emission may still put important constraint on the distribution of these ion species since most likely such emission is produced in the high-density, circumstellar regions. We performed a search to look for line emission feature in the 10 -- 15 \AA\ waveband. We did not detect any X-ray emission. Using the ObsID 7032 of 4U~1820-30 as an example, we try to determine whether such non-detection can help constrain the nature of our absorber. We find such non-detection would imply a 3$\sigma$ line flux upper limit of $6.4\times 10^{-5}\rm\ photos\ cm^{-2}s^{-1}$ for \ion{Ne}{9} $K_{\alpha}$ forbidden line at 13.69 \AA. Using our XSTAR modeling, we find the expected line flux is $2.3\times 10^{-5}\rm\ photos\ cm^{-2}s^{-1}$, lower than the 3$\sigma$ constraint from the observation. So at least in this case, non-detection of the X-ray line emission cannot provide much constraint on the distribution of the \ion{Ne}{9} ion species. However, X-ray line emission can nevertheless be an important diagnostic tool once detected.

If the \ion{Ne}{9} absorbers are intrinsic to the XRBs, photo-ionization can certainly explain some of the absorbers (see, e.g, \citealp{miller2004}). The most likely origin of the absorbing gas is perhaps the outflow from accretion disk \citep[e.g.,][]{proga2002,yuan2012b,yuan2012}. However, another interesting possibility is a jet-inflated bubble around the XRBs. Such bubbles formed when strong jet/outflow interact with the ISM \citep{weaver1977}. The ISM would be shock heated to a high temperature around $10^6 $K and later be swept up to form a thin shell of entirely HII gas. Observation of the out shell of interstellar bubble have already been performed in many systems like stellar system , stellar clusters, ultra-luminous X-ray sources, and high-mass XRBs \citep[e.g.,][]{pakull03,stevens03,fabrika04,gallo05,pakull10}. The bubble structure has also been theoretically investigated for years. The properties of the bubble mostly depend on the mass-loss rate of central target, outflow velocity and the age of the bubble \citep{{weaver1977}}. Here we simply assume an outflow rate of 10$^{-5}$ M$_{\odot}$yr$^{-1}$, bubble age of 10$^6$ yr, wind velocity of 200 km/s and ISM density of 1 cm$^{-3}$. We could estimate a \ion{Ne}{9} column density on the order of  $\sim 10^{15}$ cm$^{-2}$, by adopting a theoretical bubble density and temperature profile \citep[see equation (16) and (37) of][]{weaver1977}. The \ion{Ne}{9} column density is calculated from equation~\ref{eq:column} with solar abundance assumed. The estimate \ion{Ne}{9} column density is comparable to that measured from our XRBs sample.

Similar absorption features have been frequently detected in the X-ray spectra of background AGNs at $z=0$ (see, e.g., \citealp{nicastro2002, fang2003, anderson2010,gupta2012}). Unlike the XRBs located in the Galactic disk, the sightline toward background AGNs passes through the disk, the Galactic halo, as well as the intragroup medium in the Local Group. So potentially the comparison of the X-ray absorbers detected in the XRB and AGN spectra can offer an opportunity to study the hot gas content in the circum-galactic medium and beyond. Indeed, \citet{yao2008} performed such comparison and concluded that the halo contribution to the absorption is negligible. However,  the assumption in their paper is that the absorption detected in the XRBs is originated in the ISM. As we demonstrate in this paper, this is unlikely the case for most XRBs. The ISM for certain contributes to the observed absorption; but the measured column density can be treated only as a upper limit of the hot ISM when making comparison with those from the AGNs.

\acknowledgments
We thank Shuinai Zhang and Li Ji for helping with the XSTAR modeling. We also thank the anonymous referee for very constructive suggestions. TF was partially supported by the National Natural Science Foundation of China under grant Nos.~11243001 and 11273021, as well as the ``Fundamental Research Funds for the Central Universities" No.~2013121008. YL was supported by the National Natural Science Foundation of China under grants Nos.~11222328 and 11233006. 

{\it Facility:} \facility{Chandra}


\begin{thebibliography}{52}
\expandafter\ifx\csname natexlab\endcsname\relax\def\natexlab#1{#1}\fi

\bibitem[{{Anderson} \& {Bregman}(2010)}]{anderson2010}
{Anderson}, M.~E., \& {Bregman}, J.~N. 2010, \apj, 714, 320

\bibitem[{{Arnaud}(1996)}]{Arnaud1996}
{Arnaud}, K.~A. 1996, in Astronomical Society of the Pacific Conference Series,
  Vol. 101, Astronomical Data Analysis Software and Systems V, ed. G.~H.
  {Jacoby} \& J.~{Barnes}, 17

\bibitem[{{Bradshaw} {et~al.}(1999){Bradshaw}, {Fomalont}, \&
  {Geldzahler}}]{Bradshaw1999}
{Bradshaw}, C.~F., {Fomalont}, E.~B., \& {Geldzahler}, B.~J. 1999, \apjl, 512,
  L121

\bibitem[{{Buote} {et~al.}(2009){Buote}, {Zappacosta}, {Fang}, {Humphrey},
  {Gastaldello}, \& {Tagliaferri}}]{buote2009}
{Buote}, D.~A., {Zappacosta}, L., {Fang}, T., {et~al.} 2009, \apj, 695, 1351

\bibitem[{{Cabot} {et~al.}(2013){Cabot}, {Wang}, \& {Yao}}]{cabot2013}
{Cabot}, S.~H.~C., {Wang}, Q.~D., \& {Yao}, Y. 2013, \mnras

\bibitem[{{Cackett} {et~al.}(2008){Cackett}, {Miller}, {Raymond}, {Homan}, {van
  der Klis}, {M{\'e}ndez}, {Steeghs}, \& {Wijnands}}]{cackett2008}
{Cackett}, E.~M., {Miller}, J.~M., {Raymond}, J., {et~al.} 2008, \apj, 677,
  1233

\bibitem[{{Dickey} \& {Lockman}(1990)}]{dickeyLockman1990}
{Dickey}, J.~M., \& {Lockman}, F.~J. 1990, \araa, 28, 215

\bibitem[{{Fabrika}(2004)}]{fabrika04}
{Fabrika}, S. 2004, Astrophysics and Space Physics Reviews, 12, 1

\bibitem[{{Fang} {et~al.}(2013){Fang}, {Bullock}, \&
  {Boylan-Kolchin}}]{fang2013}
{Fang}, T., {Bullock}, J., \& {Boylan-Kolchin}, M. 2013, \apj, 762, 20

\bibitem[{{Fang} {et~al.}(2003){Fang}, {Sembach}, \& {Canizares}}]{fang2003}
{Fang}, T., {Sembach}, K.~R., \& {Canizares}, C.~R. 2003, \apjl, 586, L49

\bibitem[{{Gallo} {et~al.}(2005){Gallo}, {Fender}, {Kaiser}, {Russell},
  {Morganti}, {Oosterloo}, \& {Heinz}}]{gallo05}
{Gallo}, E., {Fender}, R., {Kaiser}, C., {et~al.} 2005, \nat, 436, 819

\bibitem[{{Galloway} {et~al.}(2008){Galloway}, {Muno}, {Hartman}, {Psaltis}, \&
  {Chakrabarty}}]{Galloway2008}
{Galloway}, D.~K., {Muno}, M.~P., {Hartman}, J.~M., {Psaltis}, D., \&
  {Chakrabarty}, D. 2008, \apjs, 179, 360

\bibitem[{{Gupta} {et~al.}(2012){Gupta}, {Mathur}, {Krongold}, {Nicastro}, \&
  {Galeazzi}}]{gupta2012}
{Gupta}, A., {Mathur}, S., {Krongold}, Y., {Nicastro}, F., \& {Galeazzi}, M.
  2012, \apjl, 756, L8

\bibitem[{{Hagihara} {et~al.}(2010){Hagihara}, {Yao}, {Yamasaki}, {Mitsuda},
  {Wang}, {Takei}, {Yoshino}, \& {McCammon}}]{hagihara2010}
{Hagihara}, T., {Yao}, Y., {Yamasaki}, N.~Y., {et~al.} 2010, \pasj, 62, 723

\bibitem[{{Homan} {et~al.}(2006){Homan}, {Wijnands}, {Kong}, {Miller}, {Rossi},
  {Belloni}, \& {Lewin}}]{Homan2006}
{Homan}, J., {Wijnands}, R., {Kong}, A., {et~al.} 2006, \mnras, 366, 235

\bibitem[{{Hynes} {et~al.}(2004){Hynes}, {Steeghs}, {Casares}, {Charles}, \&
  {O'Brien}}]{Hynes2004}
{Hynes}, R.~I., {Steeghs}, D., {Casares}, J., {Charles}, P.~A., \& {O'Brien},
  K. 2004, \apj, 609, 317

\bibitem[{{Jonker} \& {Nelemans}(2004)}]{Jonker2004}
{Jonker}, P.~G., \& {Nelemans}, G. 2004, \mnras, 354, 355

\bibitem[{{Juett} {et~al.}(2006){Juett}, {Schulz}, {Chakrabarty}, \&
  {Gorczyca}}]{juett2006}
{Juett}, A.~M., {Schulz}, N.~S., {Chakrabarty}, D., \& {Gorczyca}, T.~W. 2006,
  \apj, 648, 1066

\bibitem[{{Kallman} \& {Bautista}(2001)}]{kallman01}
{Kallman}, T., \& {Bautista}, M. 2001, \apjs, 133, 221

\bibitem[{{Kuulkers} {et~al.}(2003){Kuulkers}, {den Hartog}, {in't Zand},
  {Verbunt}, {Harris}, \& {Cocchi}}]{Kuulkers2003}
{Kuulkers}, E., {den Hartog}, P.~R., {in't Zand}, J.~J.~M., {et~al.} 2003,
  \aap, 399, 663

\bibitem[{{Liao} {et~al.}(2013){Liao}, {Zhang}, \& {Yao}}]{liao2013}
{Liao}, J.-Y., {Zhang}, S.-N., \& {Yao}, Y. 2013, \apj, 774, 116

\bibitem[{{Maller} \& {Bullock}(2004)}]{maller2004}
{Maller}, A.~H., \& {Bullock}, J.~S. 2004, \mnras, 355, 694

\bibitem[{{Mazzotta} {et~al.}(1998){Mazzotta}, {Mazzitelli}, {Colafrancesco},
  \& {Vittorio}}]{Mazzotta1998}
{Mazzotta}, P., {Mazzitelli}, G., {Colafrancesco}, S., \& {Vittorio}, N. 1998,
  \aaps, 133, 403

\bibitem[{{Miller} {et~al.}(2009){Miller}, {Cackett}, \& {Reis}}]{miller2009}
{Miller}, J.~M., {Cackett}, E.~M., \& {Reis}, R.~C. 2009, \apjl, 707, L77

\bibitem[{{Miller} {et~al.}(2008){Miller}, {Raymond}, {Reynolds}, {Fabian},
  {Kallman}, \& {Homan}}]{miller2008}
{Miller}, J.~M., {Raymond}, J., {Reynolds}, C.~S., {et~al.} 2008, \apj, 680,
  1359

\bibitem[{{Miller} {et~al.}(2004){Miller}, {Raymond}, {Fabian}, {Homan},
  {Nowak}, {Wijnands}, {van der Klis}, {Belloni}, {Tomsick}, {Smith},
  {Charles}, \& {Lewin}}]{miller2004}
{Miller}, J.~M., {Raymond}, J., {Fabian}, A.~C., {et~al.} 2004, \apj, 601, 450

\bibitem[{{Nicastro} {et~al.}(2002){Nicastro}, {Zezas}, {Drake}, {Elvis},
  {Fiore}, {Fruscione}, {Marengo}, {Mathur}, \& {Bianchi}}]{nicastro2002}
{Nicastro}, F., {Zezas}, A., {Drake}, J., {et~al.} 2002, \apj, 573, 157

\bibitem[{{Paerels} {et~al.}(2001){Paerels}, {Brinkman}, {van der Meer},
  {Kaastra}, {Kuulkers}, {den Boggende}, {Predehl}, {Drake}, {Kahn}, {Savin},
  \& {McLaughlin}}]{Paerels2001}
{Paerels}, F., {Brinkman}, A.~C., {van der Meer}, R.~L.~J., {et~al.} 2001,
  \apj, 546, 338

\bibitem[{{Paerels} \& {Kahn}(2003)}]{Paerels2003}
{Paerels}, F.~B.~S., \& {Kahn}, S.~M. 2003, \araa, 41, 291

\bibitem[{{Pakull} \& {Mirioni}(2003)}]{pakull03}
{Pakull}, M.~W., \& {Mirioni}, L. 2003, in Revista Mexicana de Astronomia y
  Astrofisica, vol. 27, Vol.~15, Revista Mexicana de Astronomia y Astrofisica
  Conference Series, ed. J.~{Arthur} \& W.~J. {Henney}, 197--199

\bibitem[{{Pakull} {et~al.}(2010){Pakull}, {Soria}, \& {Motch}}]{pakull10}
{Pakull}, M.~W., {Soria}, R., \& {Motch}, C. 2010, \nat, 466, 209

\bibitem[{{Proga} \& {Kallman}(2002)}]{proga2002}
{Proga}, D., \& {Kallman}, T.~R. 2002, \apj, 565, 455

\bibitem[{{Rasmussen} {et~al.}(2003){Rasmussen}, {Kahn}, \&
  {Paerels}}]{rasmussen2003}
{Rasmussen}, A., {Kahn}, S.~M., \& {Paerels}, F. 2003, in Astrophysics and
  Space Science Library, Vol. 281, The IGM/Galaxy Connection. The Distribution
  of Baryons at z=0, ed. J.~L. {Rosenberg} \& M.~E. {Putman}, 109

\bibitem[{{Sala} {et~al.}(2007){Sala}, {Greiner}, {Ajello}, {Bottacini}, \&
  {Haberl}}]{Sala2007}
{Sala}, G., {Greiner}, J., {Ajello}, M., {Bottacini}, E., \& {Haberl}, F. 2007,
  \aap, 473, 561

\bibitem[{{Savolainen} {et~al.}(2009){Savolainen}, {Hannikainen}, {Vilhu},
  {Paizis}, {Nevalainen}, \& {Hakala}}]{Savolainen2009}
{Savolainen}, P., {Hannikainen}, D.~C., {Vilhu}, O., {et~al.} 2009, \mnras,
  393, 569

\bibitem[{{Shaposhnikov} {et~al.}(2010){Shaposhnikov}, {Markwardt}, {Swank}, \&
  {Krimm}}]{Shaposhnikov2010}
{Shaposhnikov}, N., {Markwardt}, C., {Swank}, J., \& {Krimm}, H. 2010, \apj,
  723, 1817

\bibitem[{{Smale}(1998)}]{Smale1998}
{Smale}, A.~P. 1998, \apjl, 498, L141

\bibitem[{{Stevens} \& {Hartwell}(2003)}]{stevens03}
{Stevens}, I.~R., \& {Hartwell}, J.~M. 2003, \mnras, 339, 280

\bibitem[{{Ueda} {et~al.}(2004){Ueda}, {Murakami}, {Yamaoka}, {Dotani}, \&
  {Ebisawa}}]{ueda2004}
{Ueda}, Y., {Murakami}, H., {Yamaoka}, K., {Dotani}, T., \& {Ebisawa}, K. 2004,
  \apj, 609, 325

\bibitem[{{Verner} {et~al.}(1996){Verner}, {Verner}, \& {Ferland}}]{verner1996}
{Verner}, D.~A., {Verner}, E.~M., \& {Ferland}, G.~J. 1996, Atomic Data and
  Nuclear Data Tables, 64, 1

\bibitem[{{Weaver} {et~al.}(1977){Weaver}, {McCray}, {Castor}, {Shapiro}, \&
  {Moore}}]{weaver1977}
{Weaver}, R., {McCray}, R., {Castor}, J., {Shapiro}, P., \& {Moore}, R. 1977,
  \apj, 218, 377

\bibitem[{{Williams} {et~al.}(2006){Williams}, {Mathur}, \&
  {Nicastro}}]{williams2006}
{Williams}, R.~J., {Mathur}, S., \& {Nicastro}, F. 2006, \apj, 645, 179

\bibitem[{{Williams} {et~al.}(2005){Williams}, {Mathur}, {Nicastro}, {Elvis},
  {Drake}, {Fang}, {Fiore}, {Krongold}, {Wang}, \& {Yao}}]{williams2005}
{Williams}, R.~J., {Mathur}, S., {Nicastro}, F., {et~al.} 2005, \apj, 631, 856

\bibitem[{{Wilms} {et~al.}(2000){Wilms}, {Allen}, \& {McCray}}]{Wilms2000}
{Wilms}, J., {Allen}, A., \& {McCray}, R. 2000, \apj, 542, 914

\bibitem[{{Yao} {et~al.}(2008){Yao}, {Nowak}, {Wang}, {Schulz}, \&
  {Canizares}}]{yao2008}
{Yao}, Y., {Nowak}, M.~A., {Wang}, Q.~D., {Schulz}, N.~S., \& {Canizares},
  C.~R. 2008, \apjl, 672, L21

\bibitem[{{Yao} {et~al.}(2006){Yao}, {Schulz}, {Wang}, \& {Nowak}}]{yao2006}
{Yao}, Y., {Schulz}, N., {Wang}, Q.~D., \& {Nowak}, M. 2006, \apjl, 653, L121

\bibitem[{{Yao} {et~al.}(2009){Yao}, {Schulz}, {Gu}, {Nowak}, \&
  {Canizares}}]{yao2009}
{Yao}, Y., {Schulz}, N.~S., {Gu}, M.~F., {Nowak}, M.~A., \& {Canizares}, C.~R.
  2009, \apj, 696, 1418

\bibitem[{{Yao} \& {Wang}(2005)}]{yao2005}
{Yao}, Y., \& {Wang}, Q.~D. 2005, \apj, 624, 751

\bibitem[{{Yuan} {et~al.}(2012{\natexlab{a}}){Yuan}, {Bu}, \& {Wu}}]{yuan2012b}
{Yuan}, F., {Bu}, D., \& {Wu}, M. 2012{\natexlab{a}}, \apj, 761, 130

\bibitem[{{Yuan} {et~al.}(2012{\natexlab{b}}){Yuan}, {Wu}, \& {Bu}}]{yuan2012}
{Yuan}, F., {Wu}, M., \& {Bu}, D. 2012{\natexlab{b}}, \apj, 761, 129

\bibitem[{{Zdziarski} {et~al.}(2004){Zdziarski}, {Gierli{\'n}ski},
  {Miko{\l}ajewska}, {Wardzi{\'n}ski}, {Smith}, {Harmon}, \&
  {Kitamoto}}]{Zdziarski2004}
{Zdziarski}, A.~A., {Gierli{\'n}ski}, M., {Miko{\l}ajewska}, J., {et~al.} 2004,
  \mnras, 351, 791

\bibitem[{{Zurita} {et~al.}(2008){Zurita}, {Durant}, {Torres}, {Shahbaz},
  {Casares}, \& {Steeghs}}]{Zurita2008}
{Zurita}, C., {Durant}, M., {Torres}, M.~A.~P., {et~al.} 2008, \apj, 681, 1458

\end{thebibliography}

\end{document}